\documentclass[preprint]{aastex631}

\usepackage[T1]{fontenc}

\newcommand{\sfrac}[2]{\mathchoice%
  {\kern0em\raise.5ex\hbox{\the\scriptfont0 #1}\kern-.15em/
    \kern-.15em\lower.25ex\hbox{\the\scriptfont0 #2}}
  {\kern0em\raise.5ex\hbox{\the\scriptfont0 #1}\kern-.15em/
    \kern-.15em\lower.25ex\hbox{\the\scriptfont0 #2}}
  {\kern0em\raise.5ex\hbox{\the\scriptscriptfont0 #1}\kern-.2em/
    \kern-.15em\lower.25ex\hbox{\the\scriptscriptfont0 #2}} {#1\!/#2}}

\newcommand{\Ub}{{\bf{U}}}

\newcommand{\gb}{{\bf{g}}}

\newcommand{\omegadot}{\dot{\omega}}

\newcommand{\C}{$^{12}$C}

\newcommand{\isot}[2]{$^{#2}\mathrm{#1}$}
\newcommand{\isotm}[2]{{}^{#2}\mathrm{#1}}

\newcommand{\castro}{{\sf Castro}}
\newcommand{\amrex}{{\sf AMReX}}
\newcommand{\pynucastro}{{\sf pynucastro}}
\newcommand{\microphysics}{{\sf Microphysics}}
\newcommand{\flash}{{\sf Flash}}

\newcommand{\gcc}{\mathrm{g~cm^{-3} }}

\newcommand{\ddt}[1]{{\frac{{\partial#1}}{\partial t}}}

\usepackage{bm}

\newcommand{\Uc}{{\,\bm{\mathcal{U}}}}

\newcommand{\Rb}{{\bf R}}

\newcommand{\Adv}[1]{{\left [\boldsymbol{\mathcal{A}} \left(#1\right)\right]}}

\begin{document}
%======================================================================
% Title
%======================================================================
\title{Sensitivity of Simulations of Double Detonation Type
Ia Supernova to Integration Methodology}

\shorttitle{Double Detonation SNe Ia and Integration Methods}

\author[0000-0001-8401-030X]{Michael Zingale}
\affiliation{Department of Physics and Astronomy, Stony Brook University,
             Stony Brook, NY 11794-3800, USA}

\author[0000-0002-2839-107X]{Zhi Chen}
\affiliation{Department of Physics and Astronomy,
Stony Brook University,
Stony Brook, NY 11794-3800, USA}

\author[0000-0002-0297-0313]{Melissa Rasmussen}
\affiliation{Department of Physics and Astronomy,
Stony Brook University,
Stony Brook, NY 11794-3800, USA}

\author[0000-0002-1633-6495]{Abigail Polin}
\affiliation{The Observatories of the Carnegie Institution for Science, 813 Santa Barbara St., Pasadena, CA 91101, USA}
\affiliation{TAPIR, Walter Burke Institute for Theoretical Physics, 350-17, Caltech, Pasadena, CA 91125, USA}

\author[0000-0003-0439-4556]{Max Katz}
\affiliation{Department of Physics and Astronomy, Stony Brook University,
             Stony Brook, NY 11794-3800, USA}

\author[0000-0001-5961-1680]{Alexander Smith Clark}
\affiliation{Department of Physics and Astronomy,
Stony Brook University,
Stony Brook, NY 11794-3800, USA}

\author[0000-0003-3603-6868]{Eric T. Johnson}
\affiliation{Department of Physics and Astronomy,
Stony Brook University,
Stony Brook, NY 11794-3800, USA}

\correspondingauthor{Michael Zingale}
\email{michael.zingale@stonybrook.edu}

%======================================================================
% Abstract and Keywords
%======================================================================
\begin{abstract}
We study the coupling of hydrodynamics and reactions in simulations of the double detonation model for Type Ia supernovae.  When assessing the
convergence of simulations, the focus is usually on spatial resolution; 
however, the method of coupling the physics together as well as the
tolerances used in integrating a reaction network also play an important role.
In this paper, we explore how the choices made in both coupling and integrating the reaction
portion of a simulation (operator / Strang splitting vs.\ the simplified spectral deferred corrections method we introduced previously) influences the accuracy, efficiency, and the nucleosynthesis
of simulations of double detonations.
We find no need to limit reaction rates or reduce the
simulation timestep to the reaction timescale.  The entire simulation
methodology used here is GPU-accelerated and made freely available as part of the \castro\ simulation
code.
\end{abstract}


%======================================================================
% Introduction
%======================================================================
\section{Introduction}\label{Sec:Introduction}

Many different progenitor models for Type Ia supernovae (SNe Ia) 
are currently being explored to explain the observed spectroscopy diversity of SNe Ia explosions \citep{taubenberger:2017}. One model, the double detonation model
\citep{NOMOTO82,WOO_WEAV84}, involves a sub-Chandrasekhar mass white dwarf with an accreted He layer. A detonation in the $\mathrm{He}$ layer propagates
around the star while sending a compression wave converging toward the
center.  This compression wave ignites a carbon detonation that then
propagates outward from the center of the white dwarf, releasing enough
nuclear binding energy to gravitationally unbind the star.

Many simulations have been performed on this model, varying the initial conditions, reaction network sizes, and resolution. Recent works show
that, particularly in the case of a thin helium shell, a double detonation is viable progenitor for some SNe Ia \citep{fink:2007,mollwoosley:2013,shenbildsten:2014,glasner:2018,shen:2018,polin:2019,townsley:2019,kushnirkatz:2020,gronow:2020,gronow:2021,boos:2021,roy:2022,rivas:2022}.  These works use slightly different methods of integrating the reaction network and controlling the timestep, which we discuss in the sections below.  Our goal is to understand how sensitive a simulation of double detonations is to the details of how the reaction
network is integrated and how it is coupled to hydrodynamics.

All of these works solve the compressible Euler equations with reactive and
gravitational sources:
\begin{eqnarray}
\label{eq:euler1}
\ddt{\rho} + \nabla \cdot (\rho \Ub) &=& 0 \\
\ddt{(\rho \Ub)} + \nabla \cdot (\rho \Ub \Ub) + \nabla p &=& \rho \gb \\
\ddt{(\rho E)} + \nabla \cdot (\rho \Ub E + \Ub p) &=& \rho \Ub \cdot \gb + \rho \dot{S} \\
\label{eq:euler4}
\ddt{(\rho X_k)} + \nabla \cdot (\rho \Ub X_k) &=& \rho \omegadot_k
\end{eqnarray}
where $\rho$ is the mass density, $\Ub$ is the velocity, $E$ is the specific
total energy, $p$ is the pressure, and $X_k$ are the mass fractions
of the nuclei that react.  The gravitational acceleration,
$\gb$, is obtained by solving the Poisson equation,
\begin{equation}
\nabla^2 \Phi = 4 \pi G \rho \enskip ,
\end{equation}
(where $G$ is Newton's constant) for the gravitational potential, $\Phi$, and defining $\gb = -\nabla \Phi$.  We note that $\Phi$ is often approximated as a monopole or via a multipole expansion, which works well so long as the star
stays reasonably spherically symmetric.
The reaction network provides the species creation rate, $\omegadot_k$, and the energy generation rate, $\dot{S}$.
Finally, the system is closed via the equation of state:
\begin{eqnarray}
p &= p(\rho, e, X_k) \\
T &= T(\rho, e, X_k)
\end{eqnarray}
where $e$ is the specific internal energy, obtained as $e = E - |\Ub|^2/2$. 

This is a multiphysics system of equations and the different physical processes have different associated timescales.
Correspondingly, simulation codes use different methods to advance this system of equations.  For hydrodynamics,
explicit-in-time integration is usually used, with the timestep, $\Delta t$, restricted by the Courant limit (the precise form depends on whether we are doing
unsplit hydrodynamics,  dimensionally-split, or method of lines time-integration, see \citealt{ppmunsplit}).
%In two-dimensions, this is: \MarginPar{I thinks we can skip this! -alex}
%\begin{equation}
%\Delta t = C \min_{i,j} \left \{ \frac{\Delta x}{|u_{i,j}| + c_{i,j}}, \frac{\Delta y}{|v_{i,j}| + c_{i,j}}  \right \}
%\end{equation}
%where $\Delta x$ and $\Delta y$ are the grid width in the $x$- and $y$-directions, $u$ and $v$
%are the $x$- and $y$-components of the velocity and $c$ is the sound
%speed, and the minimum is taken over all zones on the grid.  Here, the
%Courant number, $C$ is required to be $C < 1$ for stability.  We note that
%this is the form of the stability criterion for the corner transport upwind
%formulation \citep{ppmunsplit} of hydrodynamics we use here.  Dimensionally-split
%or method-of-lines formulations will have a different, slightly more restrictive %criterion.
In contrast, reactions are usually evolved
using implicit-in-time integration, with many small steps taken to
make up the hydrodynamics timestep $\Delta t$.

A key concern in modeling astrophysical reacting flows is the coupling
of hydrodynamics and reactions.  If a large amount of energy is dumped
into a zone, then a large flow will result to carry away this energy,
requiring tight communication between the reaction and hydrodynamics
solvers.  This coupling takes various forms in astrophysical
simulation codes.  The most common method is via operator splitting:
the advection and reactive terms are treated independently, with each
process working on the result of the other.  Often, Strang-splitting
\citep{strang:1968} is used, which alternates advection and reaction
to yield second-order accuracy in time.  The primary appeal of operator
splitting is that it is easy to implement---the hydrodynamics and reaction
solvers are largely independent of one another.

When using operator splitting, the density remains constant during the reactive update, since there
is no reaction source in the mass continuity equation.  The mass
fractions, $X_k$, and energy evolve according to:
\begin{eqnarray}
\frac{dX_k}{dt} &= \omegadot_k(\rho, T, X_k) \label{eq:stranga} \\
\frac{de}{dt} &= \dot{S}(\rho, T, X_k)  \label{eq:strang_energy}\\
T &= T(\rho, e, X_k) \label{eq:strangc}
\end{eqnarray}
In the simplest approximation of operator splitting, the
evolution of the mass fractions is solved alone, without integrating
the temperature or energy, Eq.~\ref{eq:strang_energy}.  This is the form that is used, e.g., in \flash\
(\citealt{flash}, see also \citealt{townsley:2016}).  A more accurate operator splitting also includes the
energy evolution (perhaps in terms of temperature), solving the system Eqs.~\ref{eq:stranga}--\ref{eq:strangc}.  This is the default in \castro~\citep{castro,castro_joss},
and temperature evolution is also used in the works of \citet{gronow:2020,gronow:2021} following the implementation of \citet{pakmor:2012}. This evolution requires
calling the equation of state each time we evaluate the right-hand side
of the ODE system, potentially increasing the computational expense.  However, as we showed in
\citet{strang_rnaas}, this is needed to get second-order convergence.  In our \amrex-Astro \microphysics\ library \citep{microphysics},
we use C++ templating heavily to reduce the computational
expense of the EOS calls by only computing the thermodynamic quantities needed by the integrator.

Detonations can be difficult to model numerically:
the explosive energy released from reactions occurs
on a much shorter timescale than
the sound-crossing time of a zone.  Many
simulations use a simulation timestep based on the reaction
timescale.  For example, \citet{prometheus} advocated for two constraints, one that limits the change in $T$ over a timestep and another that limits the change of each mass fraction, $X_k$ over a timestep.  These can both be much more
restrictive than the hydrodynamics timescale.  Many current works instead use a reaction timestep based on the change in internal energy, $e$, over the timestep. 
 This is an option in \flash.  This energy-based reaction timestep limiter
 is used, for example, in the double detonation simulations of \citet{kushnirkatz:2020,rivas:2022}.  This helps keep the reactions
and hydrodynamics coupled, but can greatly increase the expense of the
simulation.  It is also the case that since the timestep is so small, $T$ does not change much over the simulation timestep, so evolving only the mass fractions may be reasonable.  
Artificially limiting the
rates \citep{kushnir:2013, shen:2018, kushnirkatz:2020, boos:2021} is also sometimes employed to prevent the reactions from from releasing energy faster than the sound crossing timescale of a cell.   As we will see in the simulations presented here, if the reactions explicitly know about the advection that is taking place during the burn, neither of these limiting techniques appear to be needed. 

In this paper, we focus on nucleosynthesis, exploring how reactions and hydrodynamics are coupled and the role of tolerances, the ignition of the second detonation,  and the
efficiency of the overall method.  We also look at how the results behave with resolution.  Finally, we discuss some of the difficulties encountered during the simulations.

%======================================================================
% Results
%======================================================================
\section{Simulation Setup}\label{Sec:setup}

\subsection{Numerical method}

We use the freely-available \castro\ simulation code for all the
simulations shown here.  
\castro\ solves the compressible Euler equations, Eqs. \ref{eq:euler1}--\ref{eq:euler4}, using an unsplit PPM algorithm for advection \citep{ppmunsplit,millercolella:2002,ppm}, with
the Riemann solver from \citet{colellaglaz:1985} and the general
stellar equation of state of \citet{timmes_swesty:2000}.  \castro\ also advects
internal energy together with total energy, in the dual-energy formalism introduced in \citet{bryan:1995} (see \citealt{wdmergerI} for details on the implementation).  In
all simulations, the CFL constraint is the
only timestep constraint used.  Self-gravity is done using a full
Poisson solve using geometric multigrid, with Dirichlet boundary
conditions on the domain boundary computed via a multipole expansion
with a maximum order of 6.  \castro\ uses adaptive mesh refinement (AMR) via the \amrex\ library
\citep{amrex_joss}, with 
subcycling throughout the AMR hierarchy, so finer grids are advanced at a smaller timestep than the
coarse grids.  

In \citet{castro_simple_sdc}, we introduced a method based on the
ideas of spectral deferred corrections \citep{dutt:2000, bourlioux:2003} for coupling reactions and
hydrodynamics.  We termed the method ``simplified-SDC.''  In
the simplified-SDC method, the overall
time-integration is done iteratively, with the hydrodynamics seeing an
explicit reactive source and the reaction update evolving an ODE
system that includes a piecewise-constant-in-time advective source,
$\Adv{\Uc}^{n+1/2}$,
\begin{equation}
\label{eq:simplesdc}
\ddt{\Uc} = \Rb(\Uc) + \Adv{\Uc}^{n+1/2}
\end{equation}
Here, $\Uc$ is the conserved state, $\Uc = (\rho, \rho \Ub, \rho E,
\rho X_k)^\intercal$, $\Rb(\Uc)$ are the reactive sources, and
$\Adv{\Uc}^{n+1/2}$ is an approximation to the advective update over
the timestep, as computed by the corner transport upwind formulation
of the piecewise parabolic method. Crucially, this advective
term includes an explicit reaction source term and, likewise, the reaction update,
Eq.~\ref{eq:simplesdc}, means that the reactions will ``see'' what
advection is doing as the reaction network is integrated.

The reaction system is integrated using a version of the VODE ODE
integrator \citep{vode} ported to C++.  Our modifications to this
integrator are described in \citet{castro_simple_sdc}.  We use an
analytic approximation to the Jacobian (the main approximation is that
the species derivatives of the screening function are not included)
and found that we get the best results when we disable the caching of the
Jacobian in the integrator.

In addition to the time-integration strategy, integrating the reaction
ODE system requires specifying tolerances, usually both an absolute, $\epsilon_\mathrm{abs}$, and relative tolerance, $\epsilon_\mathrm{rel}$, that are combined together into a weight of the form:
\begin{equation}
w_i = \epsilon_{\mathrm{abs},i} + \epsilon_{\mathrm{rel},i} |y_i|
\end{equation}
where the $y_i$ is one of the variables being integrated by the ODE
integrator and $i$ its the index, indicating that different tolerances
can be used for each variable.  Traditionally one set of tolerances is
used for the mass fractions and another for the energy (if it is
integrated).  We will refer to the species tolerances as $\epsilon_\mathrm{abs}(X_k)$ and $\epsilon_\mathrm{ref}(X_k)$ and the energy tolerances as $\epsilon_\mathrm{abs}(e)$ and $\epsilon_\mathrm{rel}(e)$.  Unfortunately, the tolerances used by simulations are
not normally reported in papers, and in many cases, may not be easily
controlled by a code user at runtime. Some examples can be found in publicly
available networks.  For example, the freely available version of the
{\tt aprox13} network uses $\epsilon_\mathrm{rel} = 10^{-5}$ and a
floor on the mass fractions corresponding to $\epsilon_\mathrm{abs} =
10^{-6}$.  We caution that different integrators
may use different error measures or norms so comparing the tolerance values across integrators can be difficult. 
 Since the simplified-SDC update works in terms of the
conserved variables, $\Uc$, we apply a density weighting to the absolute tolerances when comparing to the values used in Strang integration (see \citealt{castro_simple_sdc}).

All simulations use the same reaction network: the {\tt subch\_simple}
as described in \cite{zhi2023}.  This includes 22 nuclei and 94 rates
from ReacLib \citep{reaclib} and is produced with
\pynucastro\ \citep{pynucastro2}.  Importantly, we include
$\isotm{N}{14}(\alpha,\gamma)\isotm{F}{18}(\alpha,p)\isotm{Ne}{21}$
which creates protons to allow for
$\isotm{C}{12}(p,\gamma)\isotm{N}{13}(\alpha,p)\isotm{O}{16}$.  This
sequence can be faster than
$\isotm{C}{12}(\alpha,\gamma)\isotm{O}{16}$, as pointed out by
\citet{shenbildsten} and is important for getting the detonation speed
correct.  To reduce the size of the network, we have
\pynucastro\ combine some $(\alpha,p)(p,\gamma)$ and $(\alpha,\gamma)$
rates into an effective $(\alpha,\gamma)$ rate.  We also approximate
the neutron captures for $\isotm{C}{12}(\isotm{C}{12},\isotm{n}{})\isotm{Mg}{23}(\isotm{n}{},\gamma)\isotm{Mg}{24}$, $\isotm{O}{16}(\isotm{C}{12},\isotm{n}{})\isotm{S}{31}(\isotm{n}{},\gamma)\isotm{S}{32}$, $\isotm{O}{16}(\isotm{O}{16}, \isotm{n}{})\isotm{S}{31}(\isotm{n}{},\gamma)\isotm{S}{32}$, eliminating
the intermediate nucleus in each sequence.  Together these approximations make the 22 nuclei network approximate a 31 nuclei network.  Screening is provided
following the procedure in \citet{wallace:1982}, combining the
screening functions of \citet{graboske:1973,alastuey:1978,itoh:1979}.

Our choice of reaction network size falls in the middle of those reported in the literature.  \citet{townsley:2019,boos:2021} use a 55-isotope network, but do not describe whether energy / temperature is integrated together with the rates (although, based on \citealt{townsley:2016} it likely is not). 
\citet{gronow:2020,gronow:2021} use more than 30 isotopes, and use the integration method from \citet{pakmor:2012}. 
\citet{rivas:2022} use the 13-isotope {\tt aprox13} network, and likewise do not state whether energy / temperature is integrated together with the rates, although \flash\ does not usually include energy/temperature evolution with the reactions.
\citet{mollwoosley:2013} used the {\tt aprox19} network together with a table from a large network to better describe the nucleosynthesis.   While they used \castro,
the temperature evolution during the reactions was simpler then than in the current versions of \castro.  \cite{roy:2022} also use {\tt aprox19}.  
The {\tt aprox13} and {\tt aprox19} networks do not have the reaction sequences discussed in \citet{shenbildsten} and will lead to an underestimate of the He detonation speed, which in turn affects
where the compression ignites the C detonation.
None of these works discuss the integration tolerances.  Several works also disable burning in shocks (this includes \citealt{kushnirkatz:2020,gronow:2020,gronow:2021,rivas:2022} and likely others as this can be the default behavior in some codes).  We do not take that approach here, {although we briefly discuss it in appendix \ref{appendix:shocks}.  We note that allowing burning in shocks will result in a higher detonation velocity compared to disabling it.  Our focus for this work is on the ability to take large timesteps when strongly coupling reactions and hydrodynamics.  In a future paper we will look at ways of treating the burning in a shocked region in more detail.

If during the advance of a timestep an
error is generated (negative density, ODE integration fails because it cannot meet the tolerances required or takes too many steps, mass
fractions don't sum to 1, CFL constraint violated at the new time),
the step is thrown-out and retried with a smaller timestep.  This is
done on a level-by-level basis in the overall AMR subcycling
hierarchy.  This is discussed further in \citet{castro_simple_sdc}.

\subsection{Initial model}

We considered an extreme version of the double detonation scenario in
\citet{castro_simple_sdc}---a very large perturbation was applied to
drive a detonation directly in a pure carbon
white dwarf.  We showed that the Strang-splitting integration method (including energy evolution) had
difficulty with the integration unless we used tighter tolerances,
while the simplified-SDC method was well-behaved.  Here we look at a
much more realistic setup for a double detonation
and look at convergence with integration method and resolution.  Our new model uses a much smaller
perturbation that drives a He detonation around the C/O white dwarf and a
compression wave that results in the ignition of a C detonation near
the center of the star.

We will run a suite of simulations all using the same initial model,
constructed following the methodology in \citet{subchandra}.  We use
an isothermal core of C/O of $1.1~M_\odot$ with $T = 10^7~\mathrm{K}$
and a thin transition region where the temperature ramps up to
$1.75\times 10^8~\mathrm{K}$ and the composition changes to 99\%
\isot{He}{4} and 1\% \isot{N}{14}.  This He envelope is then
integrated isentropically.  The density of the transition from the
underlying CO white dwarf to the He envelope was selected to yield an
envelope mass of $0.05~M_\odot$.  The code for generating the model is
freely available \citep{initial_models}.
Our choice of white dwarf mass and
envelope mass was inspired by our previous work on modeling the convective runaway in double detonations \citep{jacobs:2016}, where we found larger envelopes and/or core masses are needed to get the localized runaway that could lead to a double detonation.

We place a small temperature perturbation in the He layer using the
same form as in \citet{castro_simple_sdc}:
\begin{equation}
  T = T_0 \left \{ 1 + X(\isotm{He}{4}) f \left [1 + \tanh(2 - \sigma) \right ] \right \}
\end{equation}
where
\begin{equation}
  \sigma = \left [ x^2 + (y - R_0)^2 \right ]^{1/2} / \lambda
\end{equation}
and
\begin{equation}
  R_0 = R_\mathrm{pert} + R_\mathrm{base}
\end{equation}
where, $R_\mathrm{base}$ is the radius at which the helium layer
begins (determined as the radius where $X(\isotm{He}{4})$ first goes beyond $0.5$) and $R_\mathrm{pert}$ is the distance above the base to put the
perturbation.  We choose $R_\mathrm{pert} = 100~\mathrm{km}$.  The
temperature is perturbed above the initial model value, denoted as
$T_0$ here.  The amplitude of the perturbation is $f = 3$ and the
scale of the perturbation is $\lambda = 12.5~\mathrm{km}$.  This is a
very small perturbation, with a peak temperature of about $1.1\times
10^9~\mathrm{K}$, spread over roughly 2 zones (in our references 20 km resolution simulations).  This
is large enough to seed the initial He detonation in the
envelope.

\begin{figure}
\centering
\plotone{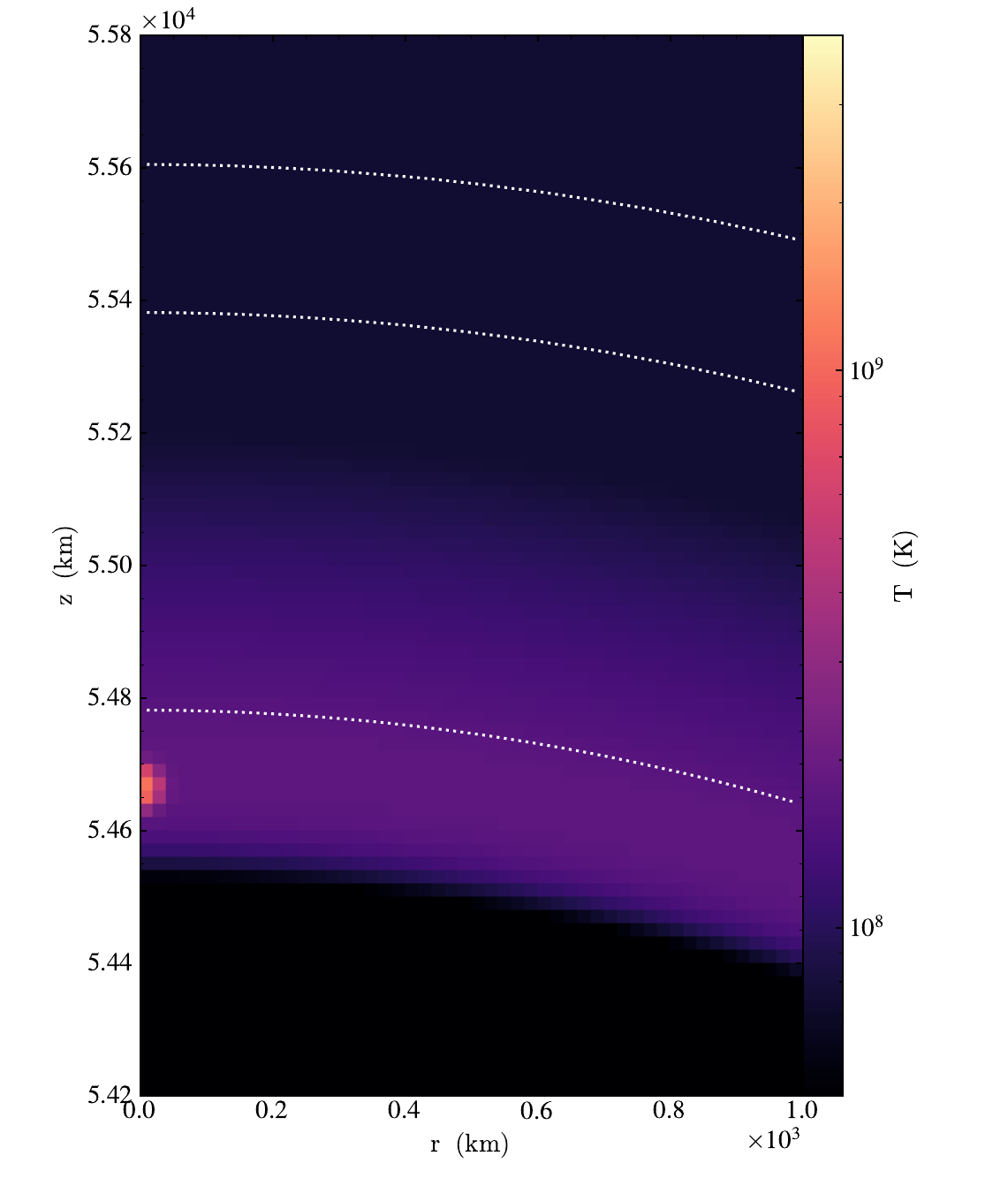}
\caption{\label{fig:initial_hotspot} Zoom-in on the initial hotspot 100 km above the He-C interface.  The dotted contours show densities
of $10^4$, $10^5$, and $10^6~\gcc$}
\end{figure}

\subsection{Suite of Runs}

All simulations use a 2D axisymmetric geometry
with a size of $5.12\times 10^9~\mathrm{cm}$ by $1.024\times
10^{10}~\mathrm{cm}$.  This is much larger than the initial size of
the star, giving it plenty of room to expand (we note that the figures
shown in the results section are all zoomed in on the star).  AMR is used, with a coarse grid of $640\times 1280$ zones.
For most simulations, 2 levels of
refinement are used (each a jump of $2\times$) giving a maximum resolution of
$20~\mathrm{km}$.  The refinement
strategy is picked to refine the star and any regions where the
temperature is greater than $10^8~\mathrm{K}$.  For the low-density regions outside of the
star, we use a sponge term on the momentum equation to prevent the
very low-density material that is not in hydrostatic equilibrium from
raining down on the star (see \citealt{eiden:2020} for a discussion of
the sponge term in \castro).

Our base simulation, which we refer to as SDC-0.2, uses the simplified-SDC integration, a CFL number
of $0.2$, and sets $\epsilon_\mathrm{rel} = \epsilon_\mathrm{abs} =
10^{-5}$ for both the species and energy.  Overall, the simulation
requires 21,784 coarse grid timesteps to evolve to $1.0~\mathrm{s}$ of
simulation time.  Table~\ref{table:simulations} summarizes the
suite of different simulations we consider.  We also show the
total amount of $\isotm{Ni}{56}$ produced at 1.0~s of simulation time for
each simulation.

\begin{deluxetable*}{lllllllll}
\tablecaption{\label{table:simulations} Parameters for our simulation suite.}
\tablehead{\colhead{name} &
\colhead{coupling} &
\colhead{CFL} &
\colhead{fine grid res} &
\colhead{$\epsilon_\mathrm{atol}(X_k)$} &
\colhead{$\epsilon_\mathrm{rtol}(X_k)$} &
\colhead{$\epsilon_\mathrm{atol}(e)$} &
\colhead{$\epsilon_\mathrm{rtol}(e)$} &
\colhead{$M_{\isotm{Ni}{56}}(t = 1.0~\mathrm{s}) / M_\odot$}}
\startdata
SDC-0.2 & simplified-SDC & 0.2 & 20 km & $10^{-5}$ &  $10^{-5}$ &  $10^{-5}$ &  $10^{-5}$ & 0.652\\
SDC-0.4 & simplified-SDC & 0.4 & 20 km & $10^{-5}$ &  $10^{-5}$ &  $10^{-5}$ &  $10^{-5}$ & 0.654 \\
\tableline
Strang-0.2 & Strang & 0.2 & 20 km & $10^{-5}$ &  $10^{-5}$ &  $10^{-5}$ &  $10^{-5}$ & 0.734 \\
Strang-0.05 & Strang & 0.05 & 20 km &$10^{-5}$ &  $10^{-5}$ &  $10^{-5}$ &  $10^{-5}$ & 0.747 \\
Strang-tol & Strang & 0.2 & 20 km & $10^{-8}$ &  $10^{-5}$ &  $10^{-5}$ &  $10^{-5}$ & 0.665 \\
\tableline
SDC-40km & simplified-SDC & 0.2 & 40 km & $10^{-5}$ &  $10^{-5}$ &  $10^{-5}$ &  $10^{-5}$ & 0.647 \\
SDC-10km & simplified-SDC & 0.2 & 10 km & $10^{-5}$ &  $10^{-5}$ &  $10^{-5}$ &  $10^{-5}$ & 0.652 \\
SDC-5km & simplified-SDC & 0.2 & 5 km & $10^{-5}$ &  $10^{-5}$ &  $10^{-5}$ &  $10^{-5}$ & 0.652 \\
\enddata
\end{deluxetable*}

The simulations are all run on the OLCF Frontier machine, using 4
nodes / 32 AMD GPUs.  The data is moved to the GPUs at the start of
the simulation and all computation is done there.  Our GPU offloading
strategy \citep{castro_gpu} takes advantage of the \amrex\ C++
parallel--for abstraction layer to be performance portable.
All of the simulation code is on GitHub\footnote{
\url{https://github.com/amrex-astro/}} and the inputs files, global
diagnostics used to make the line plots shown below, and metadata
describing the git hashes, compiler environment, and runtime
parameters for all of the simulations is available on Zenodo
at \citet{metadata}.

\section{Results}

\begin{figure*}[t]
\centering
\plotone{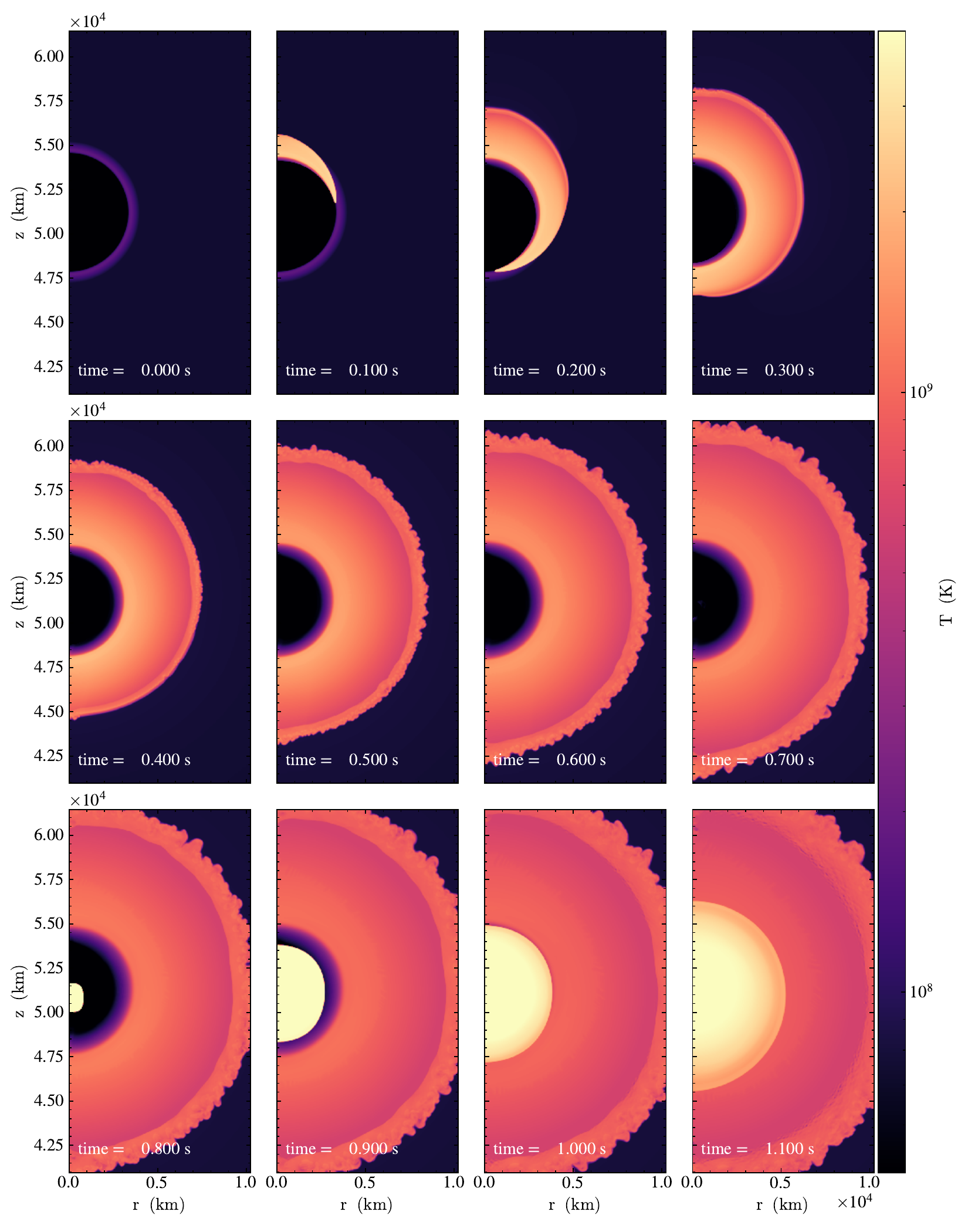}
\caption{\label{fig:temp_sequence} Time-sequence of the SDC run showing the temperature.}
\end{figure*}

Figure~\ref{fig:temp_sequence} shows the evolution of the temperature field for the base simulation.  The initial perturbation is placed at the north pole at $t=0$ and leads to a detonation in the He layer that begins to wrap around the star.  By $0.1~\mathrm{s}$ the detonation is almost halfway around the star and by $0.2~\mathrm{s}$ is almost at the south pole.  Figure~\ref{fig:lap_rho_sequence} shows a Schlieren-style plot, which
highlights density gradients (see, e.g., \citealt{svakhine:2005} a discussion of
this type of visualization).  We plot
$\log_{10}(|\rho^{-1}\nabla^2\rho|)$---this clearly shows a
compression wave launched by the early evolution of the He detonation propagating inward toward
the center of the white dwarf.  

\begin{figure*}[t]
\centering
\plotone{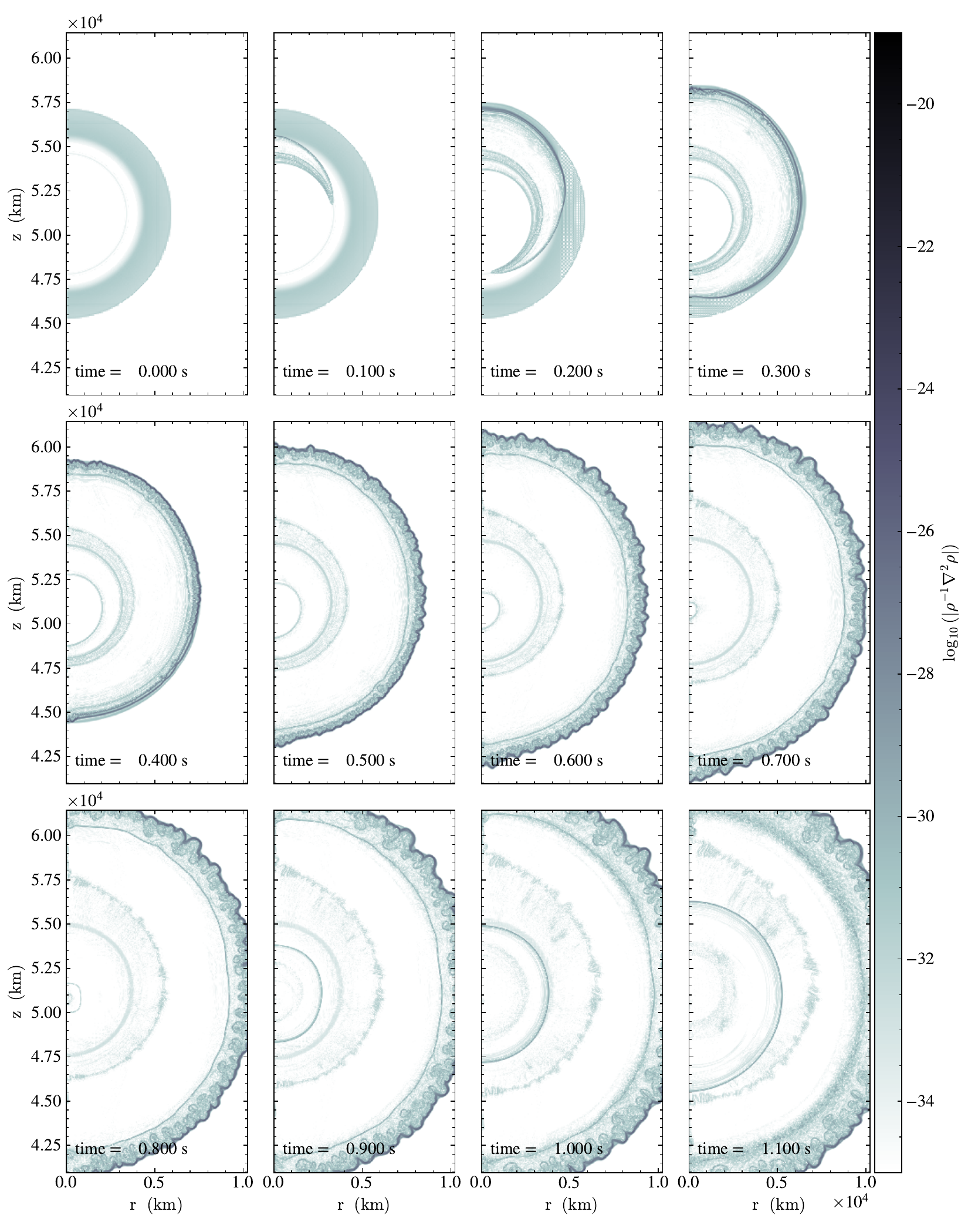}
\caption{\label{fig:lap_rho_sequence} Time-sequence of the SDC run showing the compression.}
\end{figure*}

Between $t = 0.3$ and $t = 0.7~\mathrm{s}$ the burning is confined to
the outer layer, but the compression wave continues to advance inward,
converging slightly off-center at around $t = 0.7~\mathrm{s}$.  This
highlights how the speed of the He detonation is important in
determining how off-center the C detonation ignition will be.  
A faster He detonation means the convergence happens
at higher density, making the C ignition easier.
Our
choice of network was motivated by ensuring that we capture the energy
release from He burning accurately, in particular the inclusion of the
$\isotm{C}{12}(p,\gamma)\isotm{N}{13}(\alpha,p)\isotm{O}{16}$
sequence.

In the bottom row of Figure~\ref{fig:temp_sequence}, we see
the second detonation ignites near the center of the star and
begin to propagate outward.  By $1.0~\mathrm{s}$, it has evolved
past the initial radius of the star and into the extended
ash layer left behind by the previous He detonation.

\subsection{Time-integration and Nucleosynthesis}

We now look at the details of the nucleosynthesis
and how Strang integration compares to our base simplified-SDC simulation.  We will focus on the case where the energy equation is integrated together with the species while doing Strang-splitting.  For the Strang comparisons, we will vary the CFL number and the integration tolerances.
We note that we were not able to run the case where the temperature is held fixed during the Strang burn past the C detonation ignition (for the CFL numbers we consider) without the integrator encountering problems (as we note below, the temperature became unphysically high).  This is likely why many works taking this approach use a reaction-based timestep.

\begin{figure}[t]
\centering
\plotone{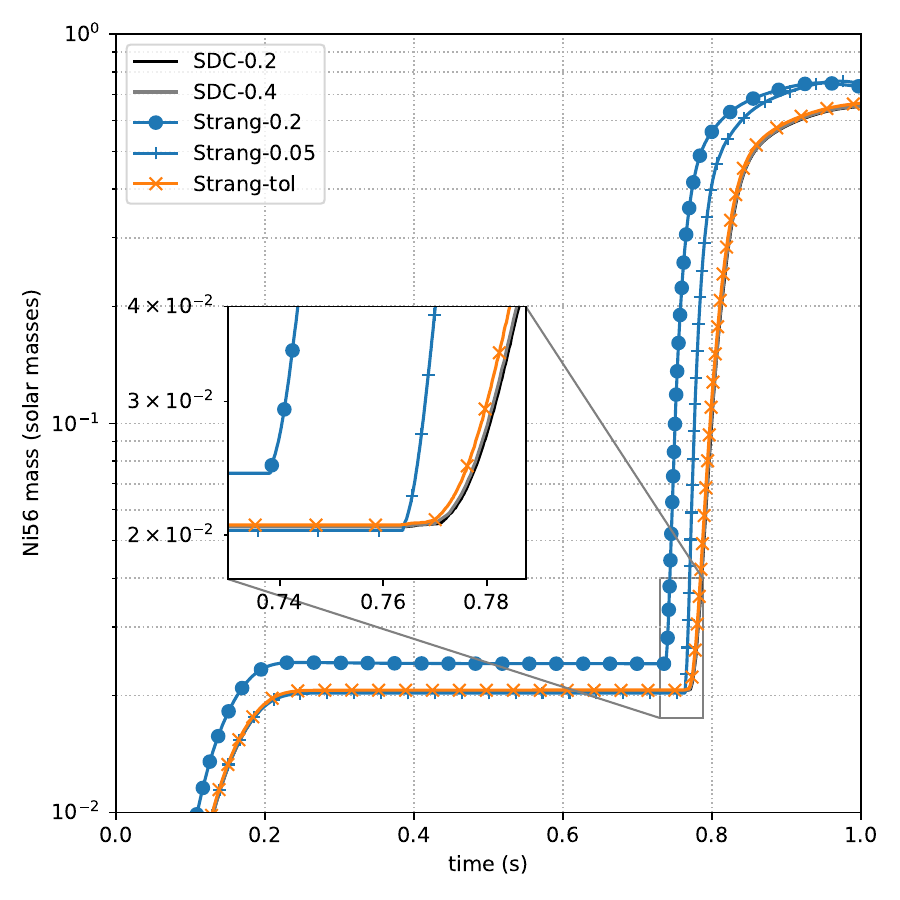}
\caption{\label{fig:ni56} \isot{Ni}{56} mass vs.\ time for the different simulations.}
\end{figure}

Figure~\ref{fig:ni56} shows the total amount of \isot{Ni}{56} as a function of time for
the suite of simulations.  The black solid line is for our SDC-0.2 simulation discussed above.  The general trend we see is that the \isot{Ni}{56} mass increases quickly until 0.2~s, corresponding to the timescale for the He detonation to wrap around the star.  About $0.02 M_\odot$ of \isot{Ni}{56} is produced during this phase.  Around 0.7~s, we see the \isot{Ni}{56} mass steeply increase, corresponding to the ignition of the C detonation and then level off at 1.0~s, corresponding to the detonation making it through the entire star.  We tabulate the \isot{Ni}{56} mass in Table~\ref{table:simulations} at 1.0~s as well.  The \isot{Ni}{56} mass fraction continues to evolve slightly beyond this point as the entire Fe-group comes into equilibrium.

The blue curves on Figure~\ref{fig:ni56} show the Strang runs with the same default integrator tolerances, but one with CFL = 0.2 (Strang-0.2) and the other with CFL = 0.05 (Strang-0.05).  We see both of them over-produce \isot{Ni}{56} at the end, but the smaller CFL run agrees with the amount of \isot{Ni}{56} produced during the He det phase.  Both lead to an earlier ignition of the C detonation.

The orange curve on Figure~\ref{fig:ni56} shows the Strang runs with the CFL = 0.2, but now different integrator tolerances.  The tolerances for
energy, unchanged: $\epsilon_\mathrm{abs}(e) = \epsilon_\mathrm{rel}(e) = 10^{-5}$, but now the simulation
labeled Strang-tol uses $\epsilon_\mathrm{abs}(X_k) = 10^{-8}$ and $\epsilon_\mathrm{rel}(X_k) = 10^{-5}$, allowing it to track lower abundances of trace species.  We see that Strang-tol matches the SDC-0.2 calculation well---the curves are nearly on top of one another.  
This agrees with what we observed with the more extreme model considered in \citet{castro_simple_sdc}, and gives us confidence that the simplified-SDC method works well.

\begin{figure}
\centering
\plotone{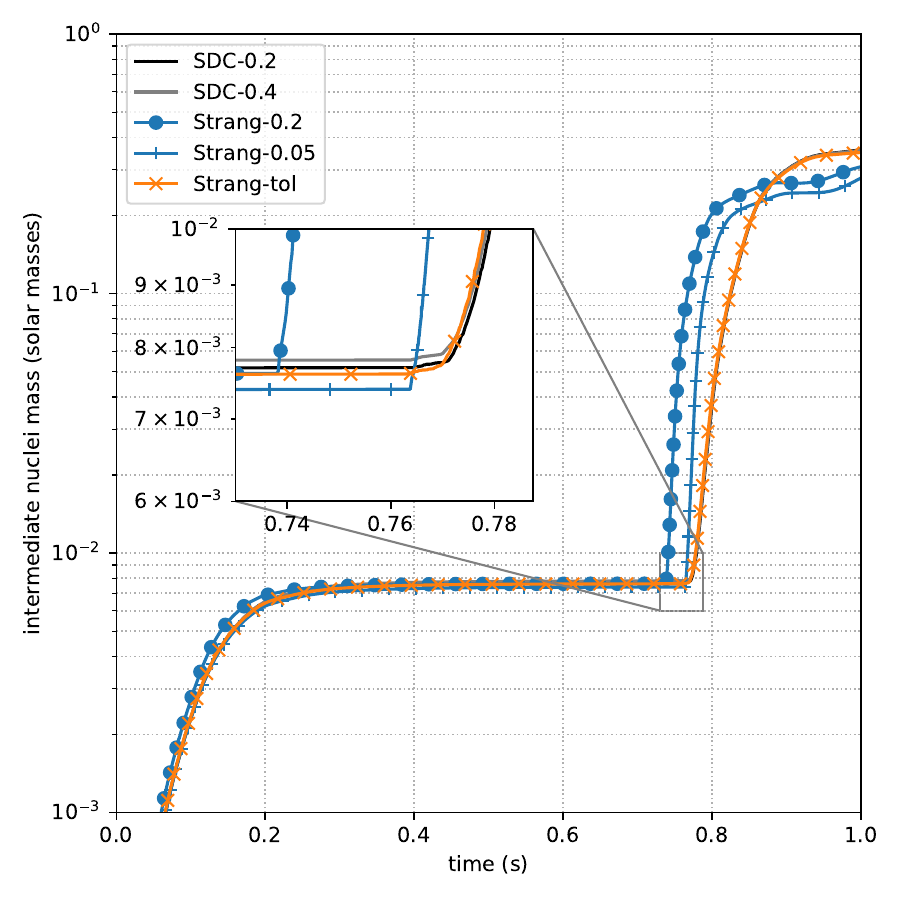}
\caption{\label{fig:subch_intermediate} Total mass of the
the nuclei from \isot{Si}{28} to \isot{Ti}{44} vs.\ time for the different simulations.}
\end{figure}

Figure~\ref{fig:subch_intermediate} shows the total mass 
of the nuclei from \isot{Si}{28} to \isot{Ti}{44} vs. time.  We see the same general trend as with the \isot{Ni}{56} plot, with the Strang-tol run agreeing with the SDC runs.  However, in this case, Strang-0.2 and Strang-0.05 underproduce these elements after the C detonation is established.

\begin{figure}
\centering
\plotone{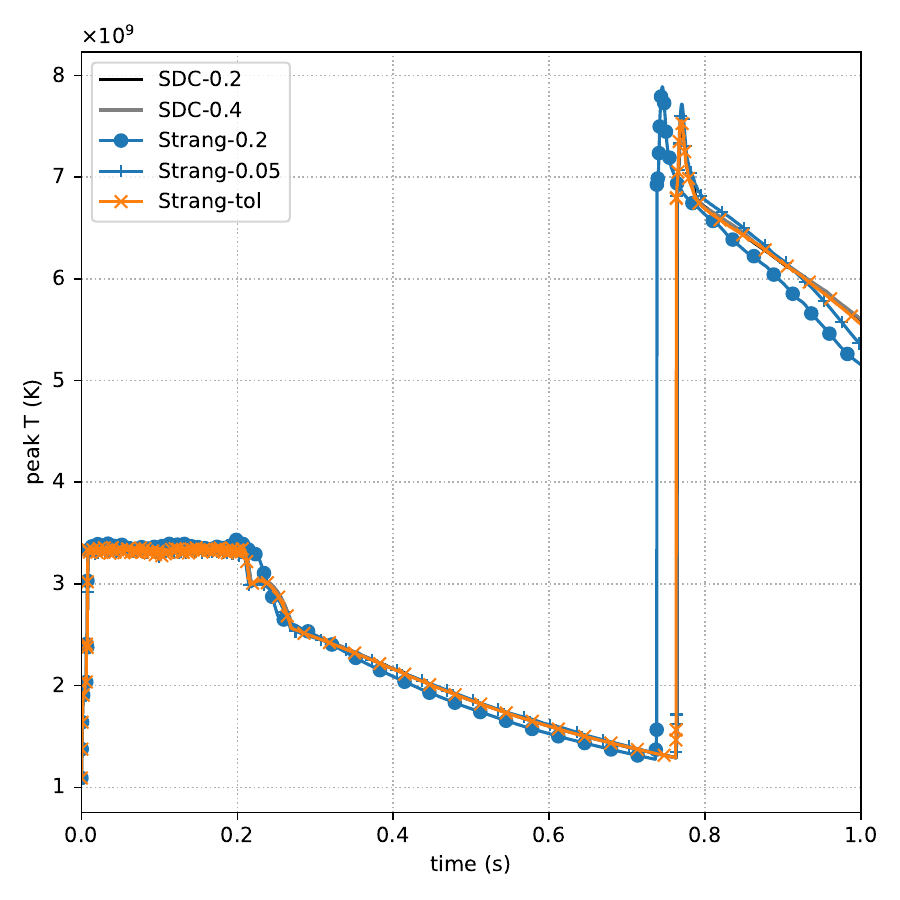}
\caption{\label{fig:subch_temp} Peak temperature vs.\ time
for the different simulations.}
\end{figure}

Figure~\ref{fig:subch_temp} shows the peak temperature vs.\ time.  From the initial hotspot, we see a quick increase to $\sim 3.3\times 10^9~\mathrm{K}$ as the He detonation propagates around the star.  The peak temperature then
begins to drop as the main burning in the He layer is over.  After $0.75~\mathrm{s}$ there is a sharp increase in the temperature to over $7\times 10^9~\mathrm{K}$---this is the ignition of the central C detonation.  We note that at this extremely high temperature, the nuclei are almost certainly in nuclear statistical equilibrium (NSE), so replacing the burning with an NSE solver would make the simulations more efficient (as done, e.g., in \citealt{ma:2013,kushnirkatz:2020}).  This will be explored in the future.  It is also interesting to
note here that the Strang-0.05 run agrees with the temperature curve for the SDC runs, so reducing the CFL number with Strang coupling does bring the temperature under control.  We also note that if 
we try to use Strang coupling without any $T$ or energy evolution, the peak temperature reaches $1.2\times 10^{10}~\mathrm{K}$ during the C detonation ignition,
and we are unable to run further (the ODE integrator encounters problems).

%To understand the difference in the \isot{Ni}{56} mass \MarginPar{should I keep this?}
%between the SDC and Strang simulations at 
%CFL = 0.2, Figure~\ref{fig:abar_sdc_strang} shows the mean molecular weight of the composition for the base SDC-0.2 and  Strang-0.2 simulations.  We see in the SDC-0.2 simulation, the highest mass ash (Ni) is confined to a thin layer on the surface of the white dwarf, while for the Strang-0.2 case, it is much more extended.  \MarginPar{zhi: will reducing the tolerance for strang (Strang-tol model) help to converge to the sdc model for this abar plot? yes, it does, which is what the curves show} The detonation in the Strang case also appears to have moved slightly faster, likely explaining why the ignition of the C detonation for this run was earlier than the base simplified-SDC simulation.

%\begin{figure}[t]
%\centering
%\plotone{sdc_strang_compare_abar_summary_plot}
%\caption{\label{fig:abar_sdc_strang} Comparison of the mean molecular weight %between SDC and Strang (both run with CFL = 0.2) at $t = 0.2~\mathrm{s}$.}
%\end{figure}

Finally, we consider what happens if we use a longer timestep with the simplied-SDC algorithm.  SDC-0.4 uses a CFL number of 0.4,
and the \isot{Ni}{56} mass is shown as the gray line in Figure~\ref{fig:ni56}.  As we see, it is virtually on top of the line corresponding to SDC-0.2, demonstrating that we remain converged when taking a larger timestep.  We note however, that the ODE integrator does have a harder time evolving the coupled system when the C detonation ignites (usually isolated to just a few zones each timestep).  As a result, at the point of the C detonation ignition, we switched the integrator to using a difference-approximation to the Jacobian \citep{lsode} instead of the analytic Jacobian,
and the simulation ran without difficulties afterwards.  This highlights however that the integrator does work hard at times, and suggests a focus on exploring different integrators or techniques to improve the robustness of the ODE integration.

\subsection{Computational Efficiency}

\begin{figure}[t]
\centering
\plotone{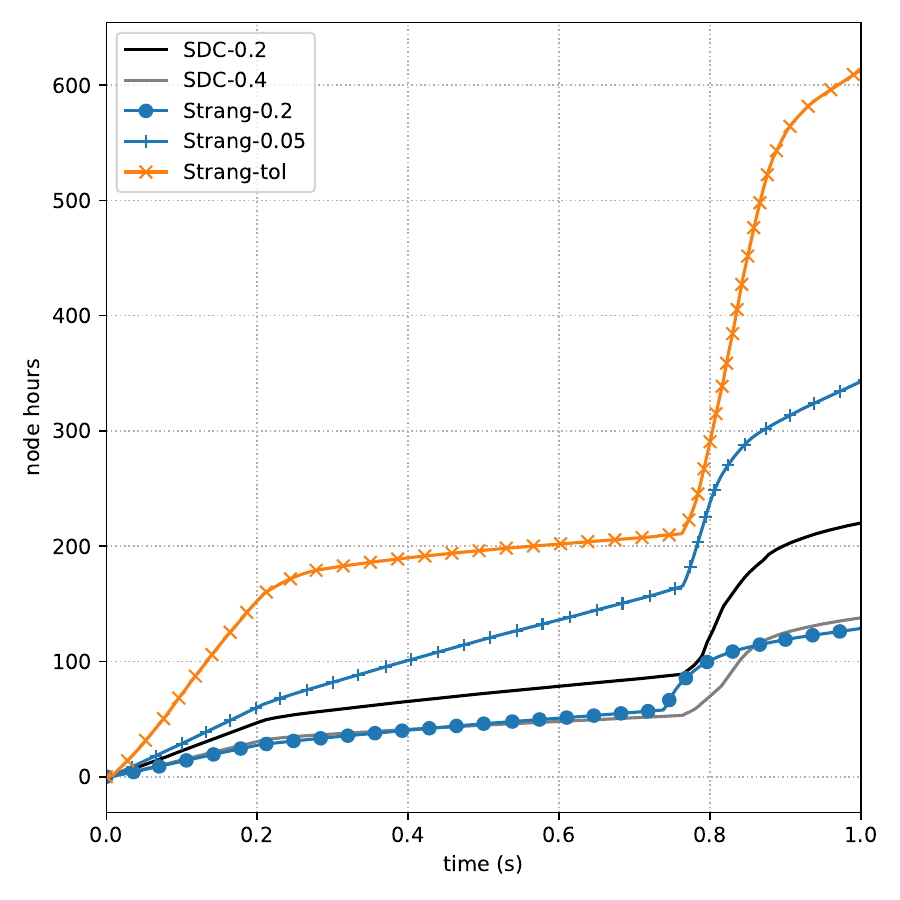}
\caption{\label{fig:cpu} node hours of GPU time for the different simulations.}
\end{figure}

We next want to ask whether the simplified-SDC method is more
expensive.
Figure~\ref{fig:cpu} shows the computational expense of the
simulations, in terms of node hours.  All simulations were run on 4
nodes (32 AMD GPUs total).  We note that we have not spent much time on
optimizing the load-balancing of these simulations, so we should consider this to be
only a guide.  \amrex\ divides the domain up into
boxes at each level.  On a level, these boxes can vary in size, but all have the same resolution.  We require 
each box to be a multiple of 32 zones on a side, with a maximum length in a dimension of 256 zones.  At the start of the simulation, there are 56 boxes on
the finest level (representing 0.77\% of the domain) and at 1 s of
evolution, there are 60 boxes covering 5.64\% of the domain.  The
figure shows that the Strang-0.2 simulation is the
cheapest, but as we saw above, this does not get the nucleosynthesis
correct.  All of the other Strang simulations are more costly than the
baseline SDC simulation.  In particular, the Strang run with the tightest tolerances, Strang-tol, that agreed with the \isot{Ni}{56} produced by SDC-0.2 is about $3\times$ more costly.       As we also see, the SDC-0.4 simulation runs very quickly---about as fast as the Strang-0.2 simulation.
This shows that not only does the
simplified-SDC simulation accurately evolves the nucleosynthesis, it
does so in a very efficient fashion.  

\subsection{Spatial resolution}

Figure~\ref{fig:res_panel} shows a time sequence of the simplified-SDC
simulation with CFL = 0.2 and our standard tolerances
($\epsilon_\mathrm{abs} = \epsilon_\mathrm{rel} = 10^{-5}$) at 4
different maximum spatial resolutions: 40 km, 20 km, 10 km, and 5 km (these are SDC-40km, our original SDC-0.2 simulation, SDC-10km, and SDC-5km in Table~\ref{table:simulations}).  For the SDC-40km
run, we simply use one fewer level of refinement compared to our SDC-0.2
simulation.  For the SDC-10km run, we add an additional level of
refinement (another jump of $2\times$, giving 3 refinement levels total) that captures only the high temperature regions (including
the initial He layer) but with the very center always refined by
adding an additional refinement tagging criteria on densities $\rho >
7.5\times 10^7~\gcc$.  For the 5 km run, we add a fourth
level of refinement (again a jump of $2\times$), but only
allow the refinement at the center to go up to this finest level of refinement.  The idea is to understand the effect of increased resolution on the C detonation ignition.  We confirmed
that for the SDC-5km simulation. the C detonation ignited in the region of the domain refined to 5~km.
The first column of the figure shows the
initial grid structure---the colors corresponding to
each level are the same for the different runs.  Recall
that the boxes at each level can vary in size, but are a multiple of 32 zones in each
dimension, with a maximum of 256 zones in any dimension.  For the SDC-10km resolution
run, we needed to switch from an analytic approximation to the Jacobian
to a numerical difference approximation once the C detonation ignited (just like with SDC-0.4), because
the evolution became too challenging for the ODE integrator otherwise.

\begin{figure*}[t]
\centering
\plotone{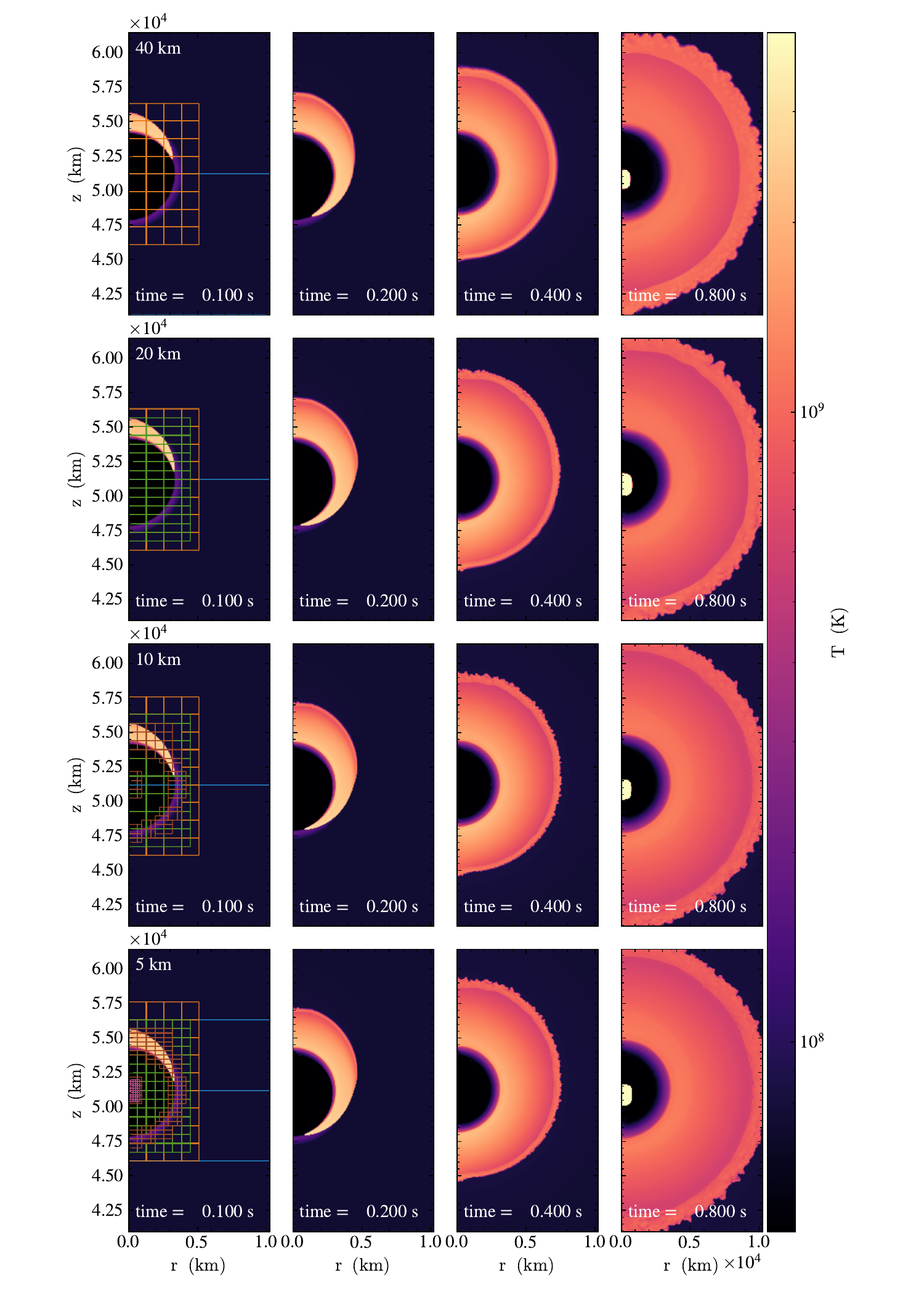}
\caption{\label{fig:res_panel} Temperature evolution for SDC simulations with CFL=0.2 at 40, 20, and 10 km.  For the earliest time
  in each resolution, the AMR grid structure is shown.}
\end{figure*}

%\begin{figure}
%\centering
%\plotone{subch_plt17035_slice}
%\caption{\label{fig:zoom-5km} Close-up of the SDC-5km simulation just after the C detonation ignited. % The grid structure is overlayed and shows that the initial detonation is completely contained in the %highest resolution region of the domain.}
%\end{figure}

%Figure~\ref{fig:zoom-5km} shows the SDC-5km run just
%around to the point when the C detonation ignites.  The
%grid structure is also overlaid, and we see that the
%initial detonation and compression wave are completely
%contained in the highest-resolution portion of the domain.

\begin{figure}[t]
\centering
\plotone{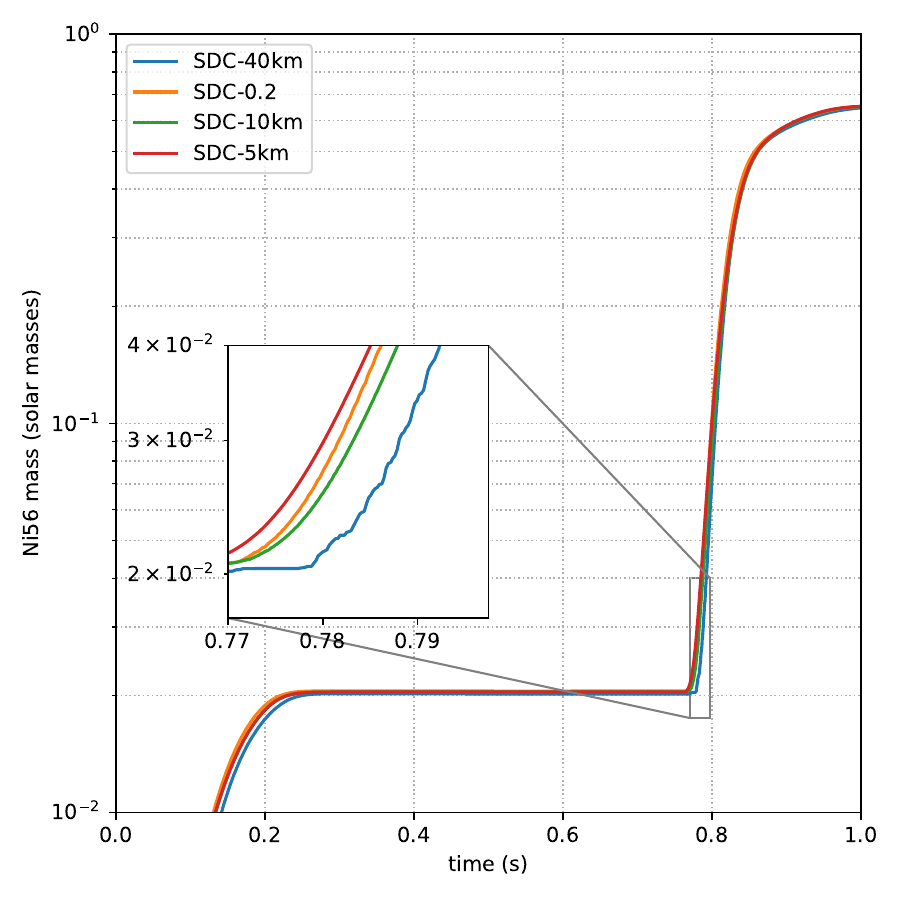}
\caption{\label{fig:res} \isot{Ni}{56} mass vs.\ time for 4 different spatial resolutions, using the simplified-SDC time integration method.}
\end{figure}

As we see from the time-evolution, the three different resolutions all
behave largely the same.  The timescale for the He detonation to
propagate around the star and the timescale at which the second C
detonation ignite match well.  Figure~\ref{fig:res} shows the
\isot{Ni}{56} mass produced from the three different simulations.
Again, the agreement is quite strong.

\begin{figure*}
\centering
\epsscale{0.9}
\plotone{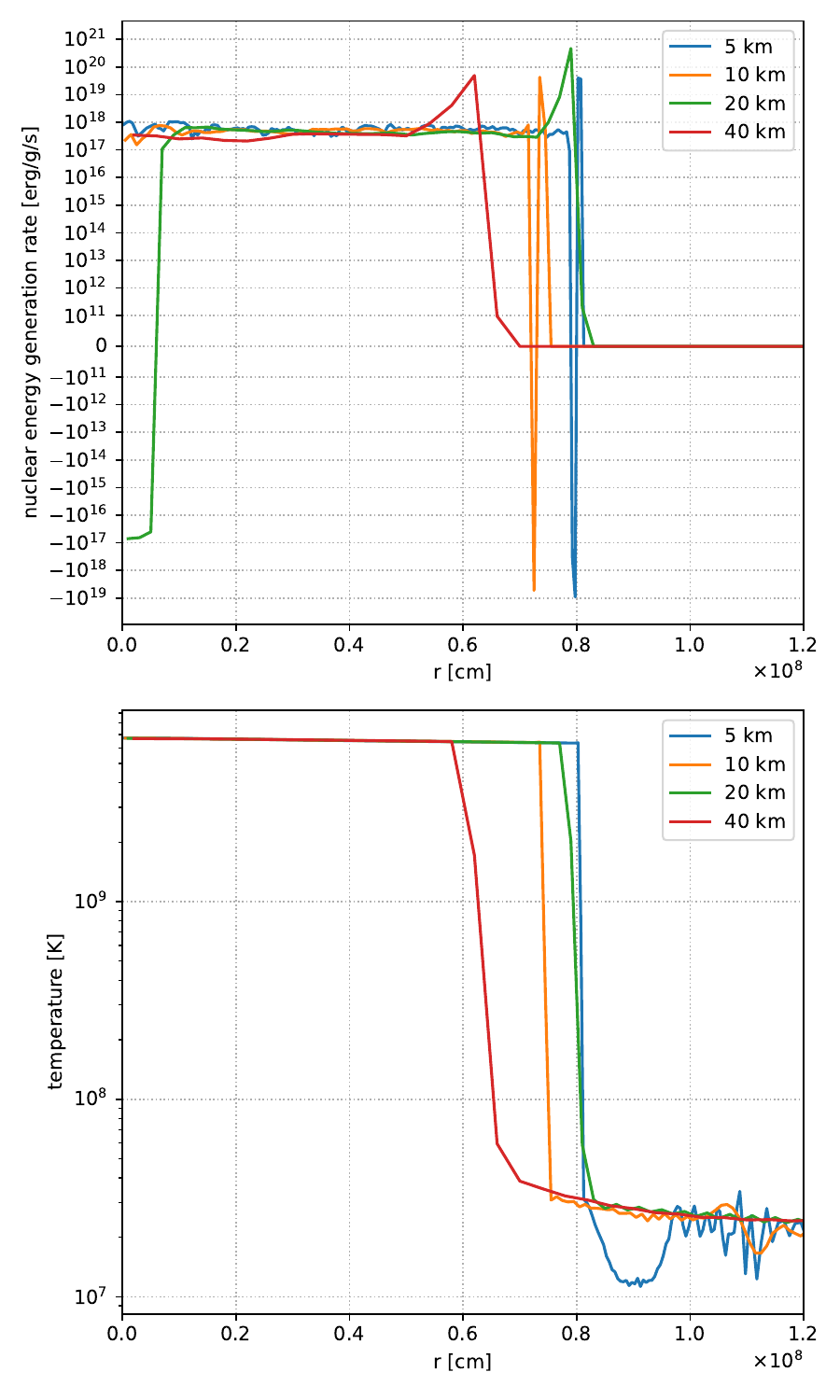}
\caption{\label{fig:det_structure} Structure of the
C detonation at $t = 0.8~\mathrm{s}$ for the 3 different resolution runs.}
\epsscale{1.0}
\end{figure*}

Figure~\ref{fig:det_structure} shows a one-dimensional
slice (along the horizontal, $r$ coordinate) through the detonation at 0.8~s.  Both the energy generation
rate and temperature are shown.  For the energy generation
rate, we see both positive and negative regions, indicating that we are attempting to find the NSE state as we evolve.  The temperature plot shows that the detonation
is very steep, particularly for the highest-resolution
simulations.  The slightly different detonation positions
arise because of the slightly different timings of the He detonation and convergence of the compression wave, which vary with resolution.

We note that these resolution results differ from the conclusions of
\citet{rivas:2022}.  There are a number of reasons why this might be
the case.  First, the initial models differ---they use a slightly lower mass white dwarf ($0.98~M_\odot$).  We also have a much broader transition between the He layer and the underlying WD than \citet{rivas:2022}.  We've noticed that the detonation speed is larger for a broader transition than a nearly discontinuous one. Second, we use a much larger reaction network here that better
captures the energetics, in particular for the initialization of the detonation, as can be seen in \cite{zhi_astronum}.  This means that the He detonation moves
faster in our simulations and the subsequent convergence of the pressure wave will be
more on-center, meaning that the ignition of the C detonation occurs
at a higher density.  Finally, we are using our new simplified-SDC time-integration, while that work used Strang-splitting.
When running at a small CFL number, spatial truncation error can build up and cause dissipation, which can only be reduced by also cutting the spatial resolution.  This is similar to the argument for temporal subcycling in adaptive mesh refinement \citep{bergercolella}.
We also have a slightly lower resolution than the finest scale used in \citet{townsley:2019}.

\section{Summary}

Our results suggest that exploring the time-integration method and
integrator tolerances are an important part of demonstrating
convergence of simulations involving explosive reacting flows in
astrophysics.  We showed that the simplified-SDC method that we introduced
in \citet{castro_simple_sdc} provides an efficient and accurate means to
model double detonation Type Ia supernovae.  In particular, for converged Ni masses, the simplified-SDC simulations required less computer time than Strang simulations. 
 And for comparable computer time, the simplified-SDC simulations were much more accurate.

Exploring the integration algorithm and tolerances, in addition to the traditional focus on resolution, was very helpful in
understanding convergence.   Our
suite of simulations show that our nucleosynthesis and dynamics appear to be converged.

The ignition of the carbon detonation remains numerically challenging
for the ODE integrator, especially with large
timesteps.  At the ignition of the C detonation, it is not uncommon for the temperature in a zone during the burn to increase from $3\times 10^9~\mathrm{K}$ to $6\times 10^9~\mathrm{K}$ during a single hydrodynamics timestep.   At these temperatures, nuclear statistical equilibrium sets in, and the integrator can spend a lot of time trying to find the equilibrium, reducing its internal timestep in the process.  The idea behind simplified-SDC coupling is that the flow should respond to a large amount of energy generated on the grid so the burning is tempered in later iterations, but this may not always be possible over a single timestep.  We will explore an alternate formulations of energy-based timestep limiters that are compatible with simplified-SDC that might help in this regime.  Sometimes, as we saw, the integrator works better in this regime if we use a numerical approximation to the Jacobian.    We are exploring heuristics to switch between analytic and numerical Jacobians based on the thermodynamics state as well as auto-differentiation methods for the Jacobian that might work more efficiently.  Additionally, we are looking at
different classes of ODE integrators that could be more efficient here.

In the near
future, we will demonstrate how to include various nuclear statistical
equilibrium approximations into the simplified-SDC formalism. 
\citet{kushnirkatz:2020} showed how enforcing NSE can help with the efficiency of the integration.  For our integration methods, where temperature varies in the reaction network as we integrate, the procedure is more complicated.  We can
begin a burn with the full network and then need to switch to NSE midway through, once the temperature increases to the point where NSE is established.  Care must be taken to do this in a second-order fashion in the simplified-SDC framework.

We will use the simplified-SDC simulation framework demonstrated here to explore the science case of the double detonation SN Ia model.    The initial state of the atmosphere is not still, but there will be a convective velocity field built up to the ignition of the detonation.  This has been modeled \citep{jacobs:2016,glasner:2018} and we can explore how the detonation propagation is affected by this state.  It is straightforward to consider larger
networks in our framework, as needed, using \pynucastro.  The follow-on goal is to add more iron-group nuclei and the weak rates that link them to our current network and explore the distribution of iron and nickel produced in these explosions.

The main three-dimensional effect to explore is multi-point ignition.  This was looked at in \citet{mollwoosley:2013}.  The question is whether the ignition of the central detonation still happens when multiple He detonations send compression waves inward.  They found that it still can occur, but this should be revisited with larger networks and our new integration methodology.  On the OLCF Frontier machine, a 20~km resolution simulation in 3D requires about $100\times$ more compute time, which translates into $\mathcal{O}(10^4)$ node hours, spread across 1024 GPUs initially (more will be needed as it refines).  This is feasible with our simulation framework.

\begin{acknowledgements}
\castro\ is open-source and freely available at
\url{https://github.com/AMReX-Astro/Castro}.  The problem setup used
here is available in the git repo as {\tt subchandra}.  The reaction network infrastructure
is contained in the AMReX-Astro Microphysics repository at \url{https://github.com/AMReX-Astro/Microphysics}.  The initial model routines
are available at \url{https://github.com/AMReX-Astro/initial_models}.

The work at Stony Brook was supported by DOE/Office of Nuclear
Physics grant DE-FG02-87ER40317.  This research used resources of the
National Energy Research Scientific Computing Center, a DOE Office of
Science User Facility supported by the Office of Science of the
U.~S.\ Department of Energy under Contract No.\ DE-AC02-05CH11231.
This research was supported by the Exascale Computing Project
(17-SC-20-SC), a collaborative effort of the U.S. Department of Energy
Office of Science and the National Nuclear Security Administration.
This research used resources of the Oak Ridge Leadership Computing
Facility at the Oak Ridge National Laboratory, which is supported by
the Office of Science of the U.S. Department of Energy under Contract
No. DE-AC05-00OR22725, awarded through the DOE INCITE program.  We
thank NVIDIA Corporation for the donation of a Titan X and Titan V GPU
through their academic grant program.  This research has made use of
NASA's Astrophysics Data System Bibliographic Services.
\end{acknowledgements}

\facilities{NERSC, OLCF}

\software{\amrex~\citep{amrex_joss},
          \castro~\citep{castro,castro_joss},
          GCC (\url{https://gcc.gnu.org/}),
          helmeos \citep{timmes_swesty:2000},
          linux (\url{https://www.kernel.org/}),
          matplotlib (\citealt{Hunter:2007}, \url{http://matplotlib.org/}),
          NetworkX \citep{networkx},
          NumPy \citep{numpy,numpy2},
          pynucastro \citep{pynucastro,pynucastro2},
          python (\url{https://www.python.org/}),
          SymPy \citep{sympy},
          valgrind \citep{valgrind},
          VODE \citep{vode},
          yt \citep{yt}}

\appendix

\section{Effect of suppressing burning in shocks}
\label{appendix:shocks}

Numerical difficulties with modeling shocks have been discussed for several decades
and stem from the fact that hydrodynamics codes smear what should be a very thin shock over several computational zones.  In a physical detonation, the burning follows behind the (infinitesimally thin) shock, and does not take place inside it \citep{prometheus}.  In the chemical combustion community, solutions to this smearing include
conservative front tracking \citep{bourlioux:1991} or using a burn fraction estimate on the subgrid scale to determine when energy should be released \citep{bdzil:2001}.

In the astrophysics community, an alternate approach is usually taken---completely disabling
burning in a zone tagged as containing a shock.  This technique goes back at least to \citet{prometheus}, where they note that a shock can be identified as any zone with a ``sufficiently large pressure gradient and negative velocity gradient'', but did not
give a threshold for the pressure gradient.  That work also focused only on 1D in a uniform density.
\citealt{papatheodore:2014} further explored the effect of burning in shocks in cellular detonation simulations.

The extension of the shock detection algorithm from 1D to multi-dimensions is not clear, and several variations appear to be in use.
The work by \citet{gronow:2020} uses a moving mesh code, but a 
structured-grid version of their criteria for a shock would be:
\begin{equation}
|\nabla P| \frac{\Delta x}{P_\mathrm{cell}} > \frac{2}{3} \qquad \nabla \cdot \Ub < 0
\end{equation}
The algorithm used in the Flash code does not appear to be published, but the source code references the AMRA code \citep{amra}.  It first detects the upstream and downstream pressures and then projects this pressure on the square of the velocity difference (only if it is negative), normalized to act like a unit vector.  It compares the gradient formed this way to the upstream pressure.

The simulations run in the main analysis allowed burning inside
of shocks, since the primary focus was on the behavior of the time-integration methodology and the effect of timestep size.  
Here we
show the behavior of the initial He detonation when burning is disabled in shocks.  
None of the above prescriptions account for the fact that in a stratified medium, the hydrostatic pressure will be unavailable to drive a shock, and therefore should be removed.  At high resolutions, this will be a small effect.  For the work shown here, we use:
\begin{equation}
\label{eq:shocktrigger}
\frac{|(\nabla p - \rho {\bf g}) \cdot \Ub_\mathrm{cell}|}{P_\mathrm{cell} |\Ub_\mathrm{cell}|} > f_\mathrm{shock} \qquad \nabla \cdot \Ub < 0
\end{equation}
to tag a zone as containing a shock.  Here, $\Ub_\mathrm{cell}$ is the velocity vector for the zone we are considering---we treat it as a unit vector and project the pressure gradient in the velocity direction since the velocity should jump across a shock and therefore roughly define the shock direction.  We treat $f_\mathrm{shock}$ as a threshold that is $\mathcal{O}(1)$.  In computing the pressure gradient without the hydrostatic term, we use a centered difference with a pressure reconstruction analogous to the well-balanced scheme described in \citet{kappeli:2016}.  If rotation was included, then the centrifugal force should also be removed. 

For the SDC algorithm, we evaluate the shock criteria once---at the start of a timestep---and then use that value throughout the entire burn.  This may be an issue, since if a shock is just entering or leaving a zone, then the shock flag will change in a zone over a single timestep.  Cutting the timestep to capture this is counterproductive, and instead we will explore interpolating the shock parameters within a zone during the burn in the future.  A similar effect will occur in Strang splitting where the shock flag may change as a result of the burning during either half timestep.

Figure~\ref{fig:shockthresh} shows the initial state of the detonation for four different values of the shock threshold parameter, $f_\mathrm{shock}$.  All simulations used 20~km resolution and CFL = 0.2. The temperature, shock flag value (where Eq.~\ref{eq:shocktrigger} tagged a shock), velocity divergence, and energy generation rate are shown.
We see for all values, there is a ``gap'' in the energy generation rate right where the shock is tagged, as expected when the burning is disabled.
Depending on the value of $f_\mathrm{shock}$, anywhere from 1--3 zones behind the detonation can be tagged as a shock.  The highest value gives a detonation speed that is roughly half the speed of the detonation that is allowed to burn everywhere.  But even the difference between $f_\mathrm{shock} = 2/3$ and $1$ is large.

\begin{figure}[t]
\centering
\includegraphics[width=\linewidth]{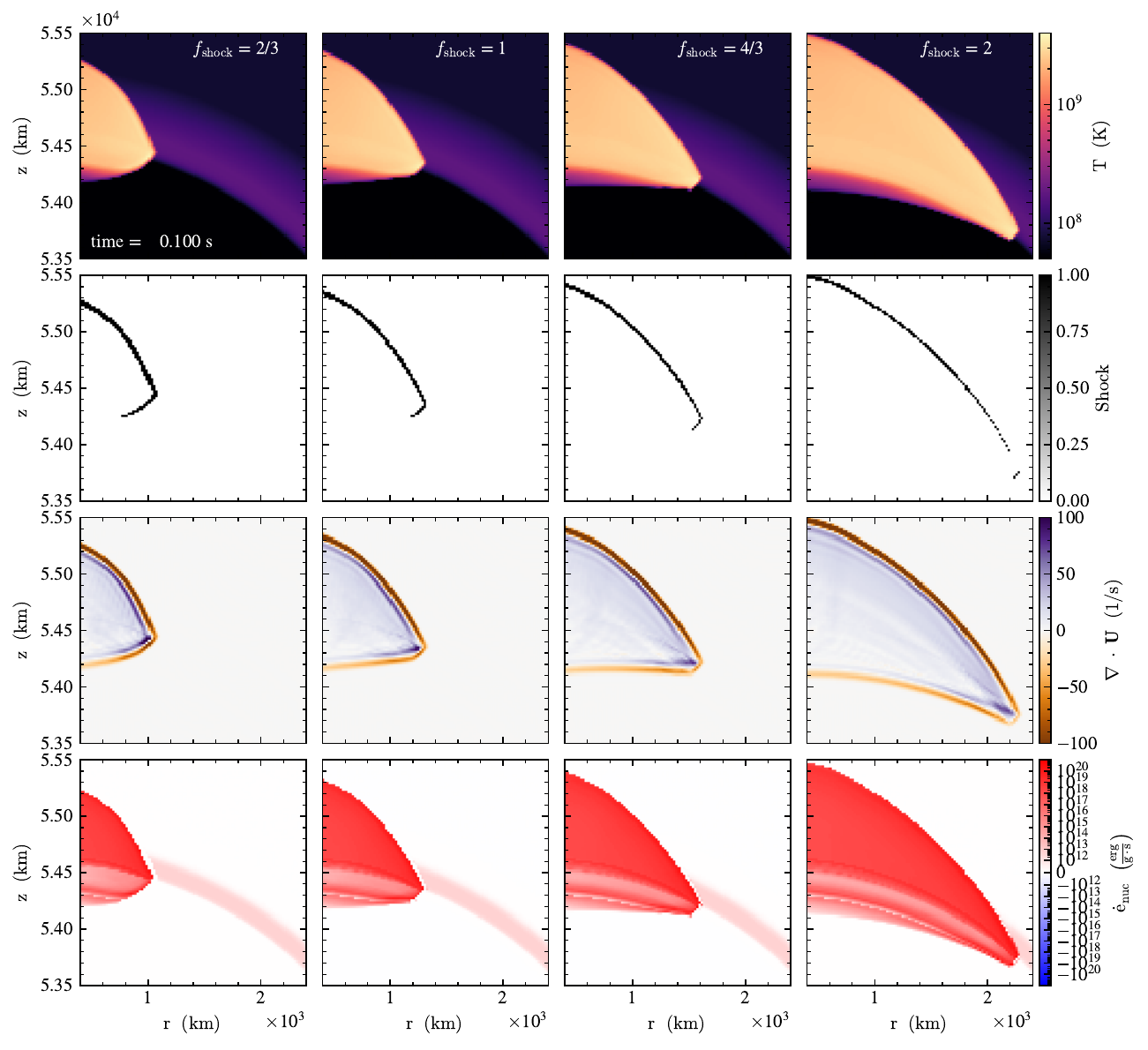}
\caption{\label{fig:shockthresh} Early evolution of the detonation with 4 different values of the shock detection threshold (columns).  Shown
are the temperature, shock detection flag, velocity divergence, and energy generation rate.}
\end{figure}

In a future paper we will explore the effect of disabling burning in a shock for detonations with the simplified-SDC method in more detail.  For this problem, we note that a side effect of a slower detonation is that the material in the He envelope ahead of the shock has more time to simmer, which can change the thermodynamic profile of the core-envelope interface and perhaps alter the evolution.

%======================================================================
% References
%======================================================================

\bibliographystyle{aasjournal}
\bibliography{ws}

\end{document}